\documentclass[twocolumn,preprintnumbers,amsmath,amssymb,superscriptaddress]{revtex4}

\usepackage{graphicx}
\usepackage{epstopdf}
\usepackage{dcolumn}
\usepackage{bm}

\begin{document}

\title{Quest for Nuclear Quadrupole Relaxation in HgBa$_2$Ca$_2$Cu$_3$O$_{8+\delta}$}

\author{Yutaka Itoh\thanks{E-mail:yitoh@cc.kyoto-su.ac.jp}}
 \affiliation{Department of Physics, Graduate School of Science, Kyoto Sangyo University, Kamigamo-Motoyama, Kita-ku, Kyoto 603-8555, Japan}
\author{Akihiro Ogawa}
 \affiliation{Chugoku Electric Power Company Inc. Technical Research Center, 3-9-1 Higashi Hiroshima, Hiroshima 739-0046, Japan}
\author{Seiji Adachi}
 \affiliation{SUSTEC, Superconducting Sensor Technology Corporation, 2-11-19 Minowa-cho, Kohoku-ku, Yokohama, Kanagawa 223-0051, Japan}

\begin{abstract}
We derived a simple reduced form from the exact solution of the mixed magnetic and quadrupolar nuclear spin--lattice relaxation function at the central transition line of a quadrupole-split NMR spectrum for the nuclear spin $I$ = 3/2.
The reduced form of the exact solution could be extended to include inhomogeneous relaxation effects.  
From the application of the reduced form to the $^{63}$Cu nuclear spin--lattice relaxation curves in the high-$T_\mathrm{c}$ superconductors of $^{63}$Cu-enriched triple-layer HgBa$_2$Ca$_2$Cu$_3$O$_{8+\delta}$, 
we estimated the predominant  magnetic relaxation rate $^{63}W_M$ and the upper limit of a weak nuclear electric quadrupole spin--lattice relaxation rate $^{63}W_Q$.   
\end{abstract}

\maketitle
\section{Introduction}
Charge stripes and short-range charge density waves in high-$T_\mathrm{c}$ copper oxide superconductors have attracted great attention~\cite{Uchida}.
Low-lying excitations associated with pairing interaction in superconductivity have been explored in multiple correlations.
NMR techniques can probe magnetic and quadrupole fluctuations through the measurements of nuclear spin relaxation times~\cite{Abragam}.
It has been a technical challenge how to detect both types of fluctuation simultaneously~\cite{Abragam,Suter,Suter2}. 

A nuclear spin--lattice relaxation time $T_1$ is experimentally estimated by the analysis of a nuclear spin--lattice relaxation curve (recovery curve).
If the recovery curve is a single exponential function with a time constant $T_1$ (spin temperature),
the relaxation rate 1/$T_1$ is expressed by the sum of a magnetic relaxation rate 1/$T_{1M}$ and quadrupolar relaxation rates 1/$T_{1Q}$'s~\cite{Abragam,Slichter}.
Then, isotopic 1/$T_1$ data serve to separate 1/$T_{1M}$ and 1/$T_{1Q}$. 
However, for unequally spacing nuclear Zeeman energy levels, the recovery curves are generally expressed by multi-exponential functions.
The time constants of the individual exponential functions are theoretically expressed by nonlinear functions of 1/$T_{1M}$ and 1/$T_{1Q}$, where transition probabilities due to quadrupole fluctuations are different from magnetic dipole transition probabilities~\cite{Suter}.            

Although exact solutions of $I$ = 3/2 nuclear spin recovery curves with both magnetic and quadrupole fluctuations have been found for unequally spacing nuclear Zeeman energy levels~\cite{Suter},
they have not been well utilized thus far. 
In Ref.~\cite{Suter}, the exact solutions of $I$ = 1 and $I$ = 3/2 are presented for the mixed magnetic and quadrupole fluctuations.

There are recent reports indicating that the theoretical recovery curve due to purely magnetic fluctuations is utilized to determine whether isotopic relaxation times of $I$ = 3/2 are magnetic or quadrupolar~\cite{Kawasaki1,Kawasaki2,Organic}.
Although the fits to the magnetic recovery curve must give us the magnetic $T_{1M}$, it has been reported that 
the experimental $T_{1M}$/$T^{\prime}_{1M}$ of two isotopes $\eta$ and $\eta^{\prime}$ disagrees with 
$(\gamma^{\prime}_n/\gamma_n)^2$ of the nuclear gyromagnetic ratios 
$\gamma_n$ and $\gamma^{\prime}_n$ and takes a close value $(Q^{\prime}_n/Q_n)^2$ of the nuclear electric quadrupole moments $Q_n$ and $Q^{\prime}_n$.  
Since the purely magnetic recovery curve assumed in anticipating the result was not justified {\it a posteriori}, 
one should analyze the data using theoretical recovery curves with both magnetic and quadrupole fluctuations from the beginning. 
Changes in the recovery curve due to quadrupole fluctuations have been overlooked in studies on the isotope dependence of $T_1$~\cite{Kawasaki1,Kawasaki2,Organic}.
Mathematical complexity may prevent us from the application of exact solutions, as is described in Ref.~\cite{Ba122}.
We need to analyze how the quadrupole relaxation process mathematically changes the purely magnetic recovery curve. 

In this paper, we studied exact theoretical $I$ = 3/2 nuclear spin--lattice relaxation curves~\cite{Suter} to simplify the mathematical treatment. 
We found a simple reduced function to estimate both magnetic and quadrupolar relaxation times and present a procedure to determine the relaxation times. 
We show the actual application of the reduced function to the high-$T_\mathrm{c}$ superconductors of $^{63}$Cu-enriched HgBa$_2$Ca$_2$Cu$_3$O$_{8+\delta}$ (Hg1223) with $T_\mathrm{c}$ = 124 and 134 K
as test samples.
The triple-layer Hg1223 is the highest-$T_\mathrm{c}$ superconductor under ambient pressure. 
Although the coexistence of antiferromagnetic ordering and superconductivity has been studied in multilayer systems~\cite{Mukuda}, the alternative effect of charge stripes has not been well explored.
We estimated the upper limit of a possible weak quadrupolar relaxation rate in Hg1223
and discussed inhomogeneous relaxation effects through stretched exponential functions introduced into the reduced forms.  

\section{Theoretical Recovery Curve}
\subsection{Exact solutions}
Let us first explain the exact solution of the rate equations (master equations) for spin $I$ = 3/2 in Ref.~\cite{Suter}. 
For a nucleus with $I$ = 3/2 in a strong static Zeeman interaction with quadrupole perturbation, three resonance lines of the $m$ = +1/2 $\leftrightarrow$ -1/2 transitions (central line) and $m$ = $\pm$3/2 $\leftrightarrow$ $\pm$1/2 transitions (satellite lines) are generally observed.
We focus on the central transition line $m$ = 1/2 $\leftrightarrow$ -1/2 in a quadruple-split NMR spectrum. 
The multi-exponential recovery curves depend on the initial condition upon spin excitation~\cite{Suter,Tsuda}. 
We consider the case of an inversion recovery technique in which a short rf inversion pulse is applied to the central transition line $m$ = 1/2 $\leftrightarrow$ -1/2 of the nuclear spin system initially in equilibrium [{\it Case I} in Ref.~\cite{Suter}]. 
The nuclear magnetization $M$($t$) is a free-induction decay and/or spin-echo signal at a time interval $t$ after an inversion pulse. 
The recovery curve is the time development of $^{}p(t) \equiv ^{}M(\infty) - ^{}M(t)$. 

The exact solution of the nuclear spin recovery curve with both magnetic and quadrupole fluctuations~\cite{Suter} is    
\begin{equation}
p(t) = a_{1c}e^{- W_A t}+a_{3c}e^{- W_B t}
\label{eqEXR}
\end{equation} 
with 
\begin{eqnarray}
W_A &=& 7W + W_1 + W_2 - b(x)W,\\
W_B &=& 7W + W_1 + W_2 + b(x)W,\\ 
a_{1c} &=& -(7+x-b(x))(1-x-b(x))/(8Wb(x)),\\
a_{3c} &=& (7+x+b(x))(1-x+b(x))/(8Wb(x)),
\label{AN}
\end{eqnarray} 
where we introduce the auxiliary parameters $x$ = ($W_1$ - $W_2$)/$W$ and $b(x)$ = $\sqrt{x^2+6x+25}$. 
The notations of $a_{1c}$, $a_{3c}$, $W$, $W_1$, and $W_2$ conform to those in Ref.~\cite{Suter}.

$a_{1c}$ and $a_{3c}$ are the coefficients of the two exponentials. 
$2W$[$\equiv$(1/$T_{1M}$)] is the magnetic nuclear spin--lattice relaxation rate with $\Delta m$ =$\pm$1. 
$2W_{1}$ and $2W_{2}$ are the nuclear electric quadrupole spin--lattice relaxation rates with $\Delta m$ =$\pm$1 and $\pm$2, respectively.
$W_1$ is a transition process different from $W$.  
The exact $a_{1c}$, $a_{3c}$, $W_A$, and $W_B$ are the analytical functions of $W$, $W_{1}$, and $W_{2}$ as in Eqs. (2)--(5). 
\begin{figure}[t]
 \begin{center} 
 \includegraphics[width=0.90\linewidth]{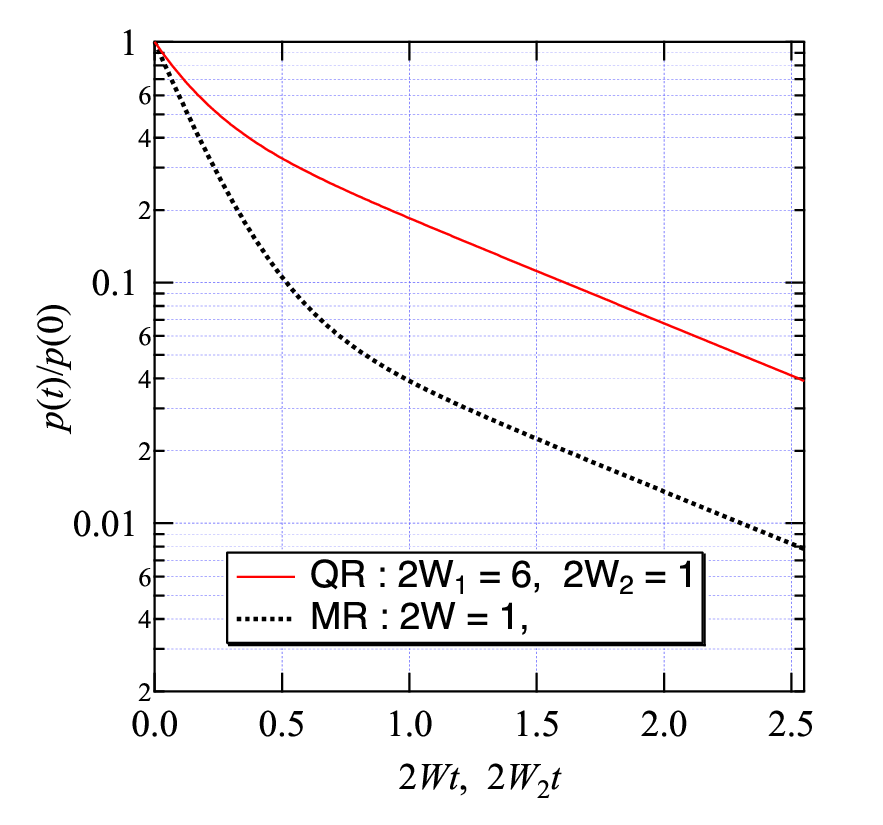}
 \end{center}
 \caption{\label{f}(Color online)
Numerical simulations of a purely magnetic relaxation curve: $p$($t$) = 0.1$e^{- t}$+0.9$e^{- 6t}$ (MR) 
and a purely quadrupolar relaxation curve: $p$($t$) = 0.5$e^{- 6t}$+0.5$e^{- t}$ (QR).
}
 \end{figure}

Equation~(\ref{eqEXR}) with $W_1$ = $W_2$ = 0 leads to the purely magnetic relaxation curve [$a_{1c}$ : $a_{3c}$ = 1 : 9 and $W_A$ : $W_B$ = 1 : 6 ]~\cite{WW},
whereas Eq.~(\ref{eqEXR}) with $W$ = 0 leads to the purely quadrupolar relaxation curve [$a_{1c}$ : $a_{3c}$ = 1 : 1 and $W_A$ : $W_B$ = $W_2$ : $W_1$ ]~\cite{Rigamonti,Borsa,Gordon}.
Figure~\ref{f} shows the numerical simulations of a purely magnetic relaxation curve of $p$($t$) = 0.1$e^{- t}$+0.9$e^{- 6t}$ [ 2$W$ = 1 ] 
and a purely quadrupolar relaxation curve of $p$($t$) = 0.5$e^{- t}$+0.5$e^{- 6t}$ [ 2$W_1$ = 6, 2$W_2$ = 1 ].
The quadruple relaxation makes the recovery curve a more single exponential.  

\subsection{Reduced forms}
Our analysis of Eq.~(\ref{eqEXR}) shows the following: 
\\
(i) In the case of $W_1$ = $W_2$, Eq.~(\ref{eqEXR}) leads to a simple reduced form as
\begin{equation}
^{}p(t) = [\frac{1}{5}e^{- 2W t}+\frac{9}{5}e^{- 12W t}]e^{- 2W_2 t}
\label{eqMQ}
\end{equation}  
for any $W$ and $W_2$. Experimentally, $W$ and $W_2$ are the fitting parameters. 
The quadrupole relaxation rates $W_1$ and $W_2$ generally are of the same order of magnitude~\cite{YM,Oba}.
Nonlinear expressions of the time constants in Eq.~(\ref{eqEXR}) lead to a linear expression of Eq.~(\ref{eqMQ}) under a reasonable condition. 
This is a primary result of this study.
\\ 

In passing, the exact solution of the recovery curve of the satellite transition line ($m$ = $\pm$3/2 $\leftrightarrow$ $\pm$1/2) is simplified to
\begin{equation}
p(t) = \frac{1}{5}e^{- 2W t - 2W_2 t}+e^{- 6W t - 4W_2 t}+\frac{4}{5}e^{- 12W t - 2W_2 t}
\label{eqSMQ}
\end{equation} 
for $W_1$ = $W_2$.
 \\     
 
\begin{figure}[t]
 \begin{center} 
 \includegraphics[width=0.90\linewidth]{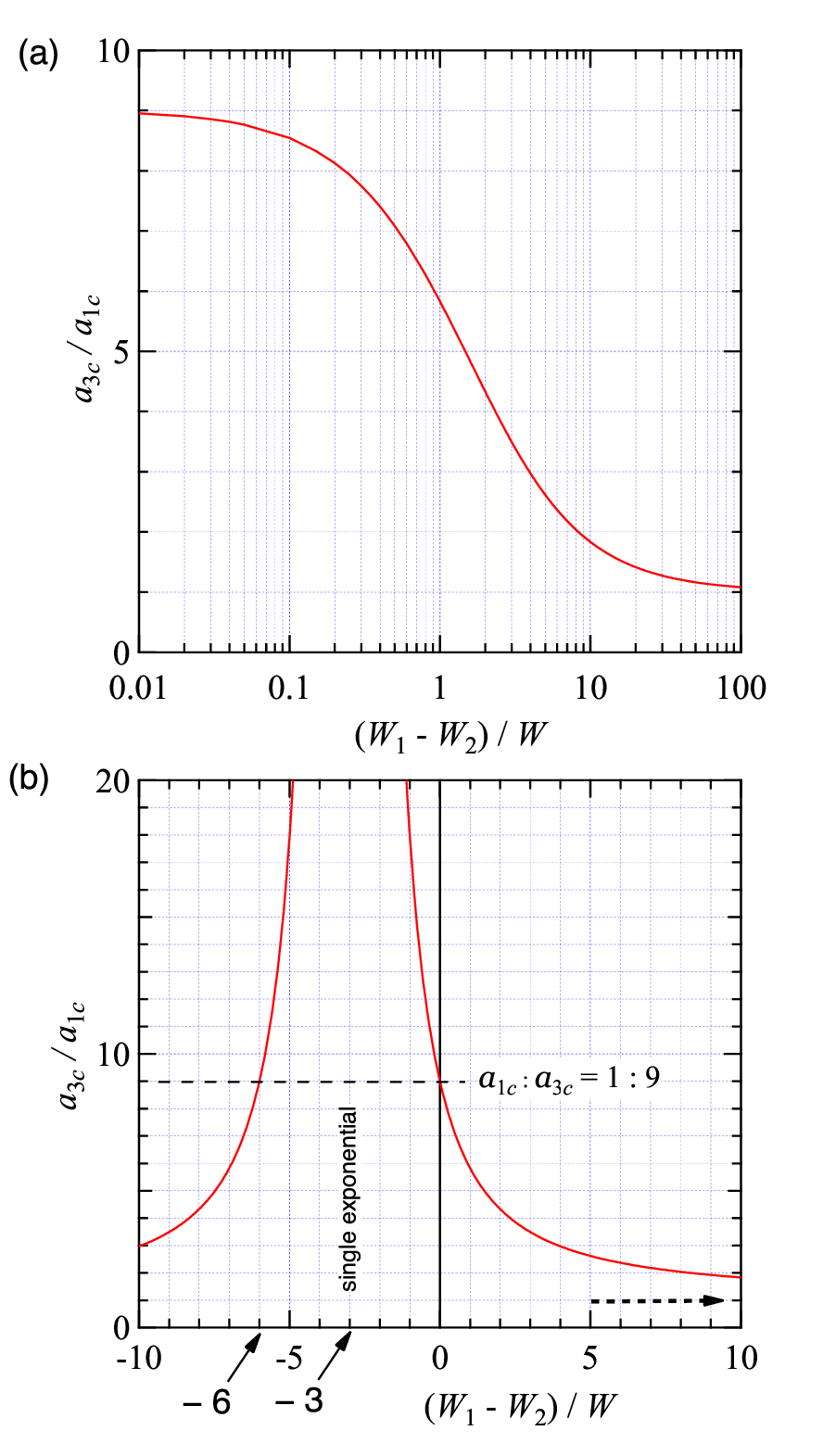}
  \hspace*{-3.0cm}
 \end{center}
 \caption{\label{SM}(Color online)
Linear-log (a) and linear (b) plots of $a_{3c}/a_{1c}$ against $x$ = $(W_1 - W_2)/W$. 
}
 \end{figure}

(ii) For $W_1 \neq W_2$, we observed that the ratio $a_{3c}/a_{1c}$ is a key to determining $W$, $W_1$, and $W_2$ in turn. 
The analytical expression of $a_{3c}/a_{1c}$ as a function of $x$ = $(W_1 - W_2)/W$ is   
\begin{equation}
\frac{a_{3c}}{a_{1c}} = -\frac{(7+x+b(x))(1-x+b(x))}{(7+x-b(x))(1-x-b(x))}.  \\
\label{AR}
\end{equation}
In Fig.~\ref{SM}, $a_{3c}/a_{1c}$ in Eq.~(\ref{AR}) is plotted against $x$ = $(W_1 - W_2)/W$ in semi-log plots (a) and linear plots (b). 

In Fig.~\ref{SM}(a), $a_{3c}/a_{1c}$ = 9 for $x$ = 0 [ $W_1$ = $W_2$ ] is the same as the purely magnetic relaxation,
and $a_{3c}/a_{1c}$ $\rightarrow$ 1 for $x$ $\gg$ 1 [ ($W_1$ - $W_2$) $\gg$ $W$ ] is the same as the purely quadrupolar relaxation. 
There is a one-to-one correspondence between $a_{3c}/a_{1c}$ and $x$ = $(W_1 - W_2)/W$ for $x >$ $-$ 3.   

In Fig.~\ref{SM}(b), for $x$ = $-$ 3 and then $a_{1c}$ = 0, the recovery curve is a single exponential function:
\begin{equation}
^{}p(t) = 2e^{- (8W + 2W_2) t}.
\label{EXP}
\end{equation}
For $-$ 6 $< x < -$ 1 and then $a_{1c} \ll a_{3c}$, the recovery curve is nearly a single exponential function.
The isotopic measurements of the single relaxation rate (8$W$ + 2$W_2$) need to separate $W$ and $W_2$.  
For $x$ = $-$ 6, one finds $a_{3c}/a_{1c}$ = 9 but $W_A/W_B$ = (8$W$+2$W_1$)/(18$W$+2$W_1$) $\geq$ 4/9 $>$ 1/6. 
That is, there is no solution with $x \ne$ 0 identical to the purely magnetic recovery curve. 
The strong $W_2$ process makes the recovery curve a more single exponential.  

From Eq.~(\ref{AR}), we found the analytical expression of $(W_1 - W_2)/W$ as a function of $a_{3c}/a_{1c}$:
\begin{equation}
\frac{W_1 - W_2}{W} = - 3 \pm 8 \frac{\sqrt{(a_{3c}/a_{1c})}}{(a_{3c}/a_{1c}) -1 }. 
\label{EX}
\end{equation}
Some assumptions may have to be introduced to determine which is appropriate, $+$ or $-$.  
  
Experimentally, $a_{1c}$, $a_{3c}$, $W_A$, and $W_B$ are the fitting parameters in the double exponential function of Eq.~(\ref{eqEXR}).   
Then, one can estimate the value of $x$ = $(W_1 - W_2)/W$ from the experimental ratio $a_{3c}/a_{1c}$ through Eq.~(\ref{EX}) and
then the value of $b(x)$ = $\sqrt{x^2+6x+25}$. 
The procedure to estimate $W$, $W_1$, and $W_2$ in turn is 
\begin{eqnarray}
2W &=& \frac{W_B - W_A}{b}, \\
W_1 + W_2 &=& \frac{W_A + W_B}{2} - 7W, \\
W_1 - W_2 &=& xW,
\label{Qr}
\end{eqnarray}
and 
\begin{eqnarray} 
2W_1 &=& \frac{7+b-x}{2b}W_A - \frac{7-b-x}{2b}W_B, \\
2W_2 &=& \frac{7+b+x}{2b}W_A - \frac{7-b+x}{2b}W_B.
\label{Qr}
\end{eqnarray} 
\begin{figure}[t]
 \begin{center} 
 \includegraphics[width=0.90\linewidth]{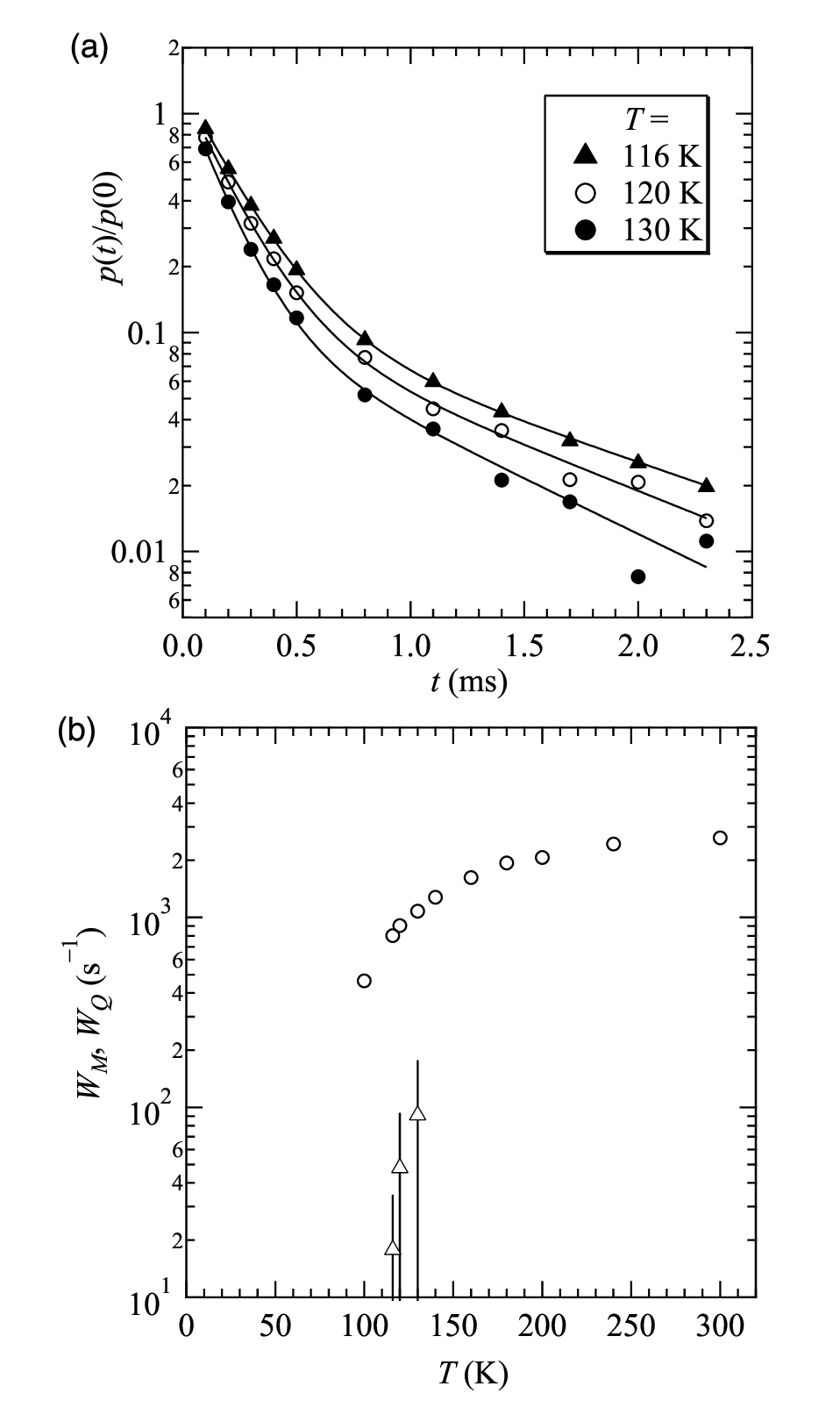}
 \end{center}
 \caption{\label{L}
(a) Recovery curves $^{63}p$($t$) of the $^{63}$Cu nuclear spin-echo signals for underdoped Hg1223 ($T_\mathrm{c}$ = 124 K) at an external magnetic field $B_0$ of 7.5 T along the $c$-axis ($B_0 \parallel c$). 
Solid curves are the least-squares fitting results obtained using Eq.~(\ref{MQ}). 
(b) Magnetic $^{63}$Cu nuclear spin--lattice relaxation rate $^{63}W_M$ (open circles) and nuclear electric quadrupole spin--lattice relaxation rate $^{63}W_Q$ (open triangles) versus $T$ in semi-log plots for underdoped Hg1223 with $B_0 \parallel c$. 
The central transition lines of the inner plane $^{63}$Cu(1) and the outer plane $^{63}$ Cu(2) are indistinguishable for the present underdoped Hg1223 with $B_0 \parallel c$. 
 }
 \end{figure}    

\section{Experimental Test} 
We applied Eq.~(\ref{eqMQ}) to the actual $^{63}$Cu nuclear spin recovery curves in the triple-layer Hg1223 to determine whether Eq.~(\ref{eqMQ}) achieves a rapid least-squares fitting.   
The NMR experiments were performed for the magnetically $c$-axis-aligned samples of $^{63}$Cu-enriched Hg1223 (underdoped $T_\mathrm{c}$ = 124 K and optimally doped $T_\mathrm{c}$ = 134 K),
which have been reported in Ref.~\cite{Itoh}.

Figure~\ref{L}(a) shows the experimental recovery curves $^{63}p$($t$) of the $^{63}$Cu nuclear spin-echo signals for the underdoped Hg1223 ($T_\mathrm{c}$ = 124 K) at an external magnetic field $B_0$ of 7.5 T along the $c$-axis ($B_0 \parallel c$).
The solid curves are the least-squares fitting results obtained using the function of Eq.~(\ref{eqMQ}),  
\begin{equation}
^{}p(t) = p(0)[0.1e^{- W_M t}+0.9e^{- 6W_M t}]e^{- W_Q t},
\label{MQ}
\end{equation}
where $p$(0), $W_M$($\equiv$2$W$), and $W_Q$($\equiv$2$W_2$) are the fitting parameters. 

Figure~\ref{L}(b) shows $^{63}W_M$ and $^{63}W_Q$ versus temperature $T$ for the underdoped Hg1223 with $B_0 \parallel c$.
We estimated $^{63}W_Q$ ($\leq$ 10$^{-2}$ s$^{-1}$) with large experimental uncertainty above 140 K and below 100 K and $^{63}W_M$ = 4.6$\times$10$^2$ - 2.6$\times$10$^3$ s$^{-1}$ at $T$ = 100--300 K. 
The spin pseudogap behavior in $^{63}W_M/T$ was still observed at $T^*$ = 190 K~\cite{Itoh}.       
The finite $^{63}W_Q$ around 110 $< T <$ 140 K reminds us of that of the $^{17}$O nuclear quadrupole relaxation rate $^{17}W_2$ in underdoped YBa$_2$Cu$_4$O$_8$ ($T_\mathrm{c}$ = 82 K)~\cite{Suter2}.

For the optimally doped Hg1223 with $B_0\parallel c$ and for the underdoped Hg1223 with $B_0\perp c$, the NMR lines of the inner-plane $^{63}$Cu(1) and outer-plane $^{63}$Cu(2) sites were distinguishable~\cite{Itoh}. 
We obtained $^{63}W_Q$ $\leq$ 1$\times$10$^{-2}$ s$^{-1}$ (upper limit) at both $^{63}$Cu(1) and $^{63}$Cu(2) sites in both compounds. 

Equation~(\ref{MQ}) is used to estimate both $^{63}W_M$ and $^{63}W_Q$ separately from the central transition line. 
Experimental resolution on the recovery curve data is crucial to determine $^{63}W_M$ and $^{63}W_Q$.  
Further precise measurements of the recovery curves are required to confirm $^{63}W_Q$, because the deviation of the actual recovery curve from the purely magnetic function is small and the values of $^{63}W_Q$ are comparable to the experimental uncertainty.

\section{Distribution of 1/$T_1$}
\begin{figure}[b]
 \begin{center} 
 \includegraphics[width=0.90\linewidth]{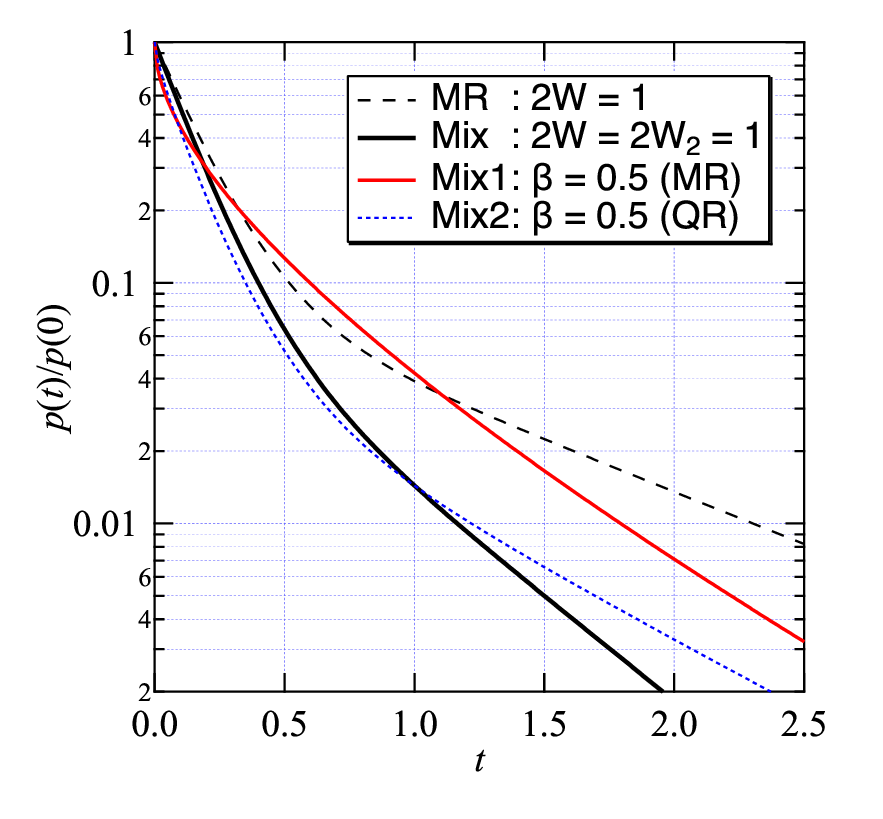}
 \end{center}
 \caption{\label{Mix}(Color online)
Numerical simulations of a purely magnetic relaxation curve: $p$($t$) = 0.1$e^{- t}$+0.9$e^{- 6t}$ (MR: dashed curve), 
a mixed magnetic and quadrupolar relaxation curve: $p$($t$) = [ 0.1$e^{- t}$+0.9$e^{- 6t}$ ]$e^{- t}$ (Mix: solid curve),
a stretched magnetic mixed relaxation curve: $p$($t$) = [ 0.1$e^{- \sqrt{t}}$+0.9$e^{- \sqrt{6t}}$ ]$e^{- t}$ (Mix1: red curve),
and a stretched quadrupolar mixed relaxation curve: $p$($t$) = [ 0.1$e^{- t}$+0.9$e^{- 6t}$ ]$e^{- \sqrt{t}}$ (Mix2: blue dotted curve).
}
 \end{figure}
For an inhomogeneous relaxation process, stretched exponential functions can be introduced into the exact solution.
The distribution averages on $W_M$ and $W_Q$ can be taken independently in Eq.~(\ref{MQ}), because they are represented by Laplace transformations for heavy-tailed distribution functions~\cite{ItLSCO,DCJ,str,JK}.    
From the distribution average on $W_M$, the reduced form of Eq.~(\ref{MQ}) can be transformed into the stretched exponential functions as
\begin{equation}
^{}p(t) = p(0)[0.1e^{- (W_M t)^{\beta}}+0.9e^{- (6W_M t)^{\beta}}]e^{- W_Q t},
\label{sMQ}
\end{equation}
where $p$(0), $W_M$, $W_Q$, and $\beta$ are the fitting parameters. 
The variable exponent $\beta$ characterizes a cut-off value of low relaxation rates in the distribution function and 
1/$\beta$ characterizes the average of the relaxation time~\cite{DCJ}. 
From the distribution average on $W_Q$, Eq.~(\ref{sMQ}) can be transformed into 
\begin{equation}
^{}p(t) = p(0)[0.1e^{- (W_M t)^{\beta}}+0.9e^{- (6W_M t)^{\beta}}]e^{- (W_Q t)^{\beta_Q}},
\label{MQs}
\end{equation}
where $p$(0), $W_M$, $W_Q$, $\beta$, and $\beta_Q$ are the fitting parameters.  
\begin{figure}[t]
 \begin{center} 
 \includegraphics[width=0.90\linewidth]{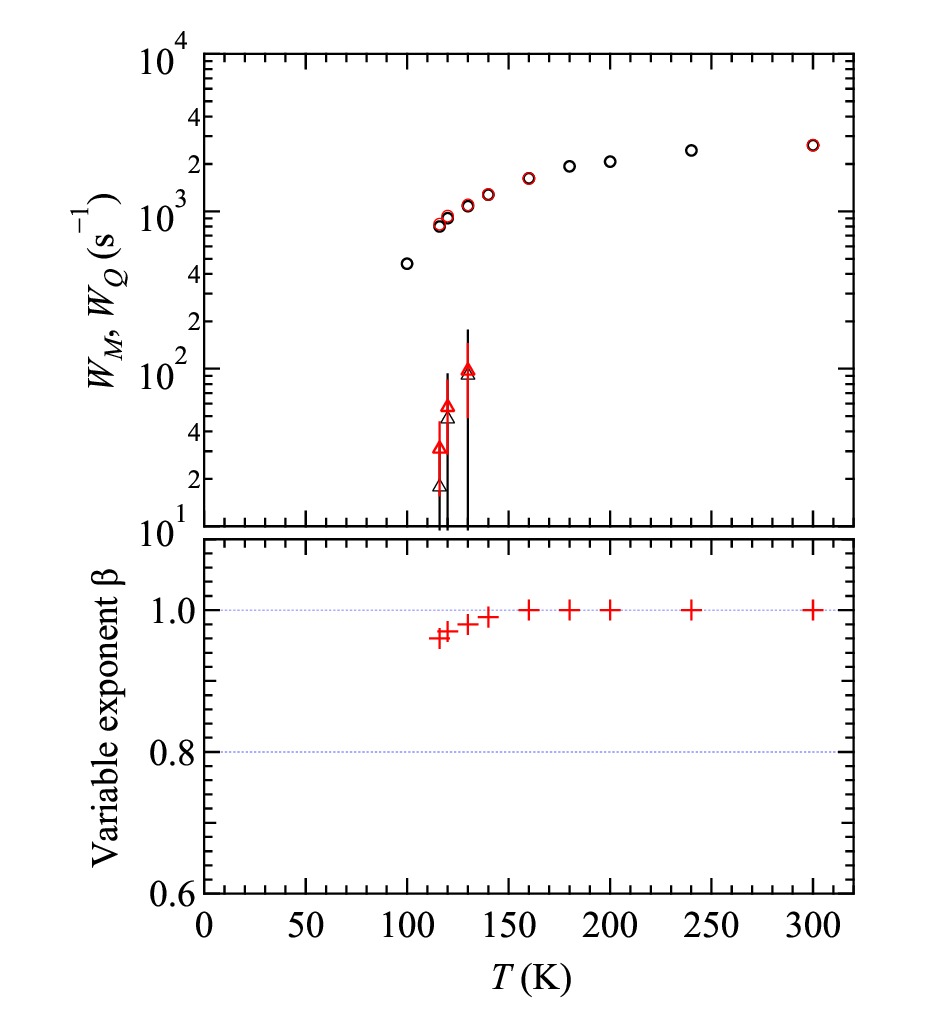}
 \vspace*{+1.0cm}
 \end{center}
 \caption{\label{TT}(Color online)
Red symbols are the least-squares fitting results obtained using Eq.~(\ref{sMQ}) for underdoped Hg1223 with $B_0 \parallel c$. 
Upper panel: $^{63}W_M$ (red open circles) and $^{63}W_Q$ (red open triangles) versus $T$ in semi-log plots. 
Lower panel: variable exponent $\beta$ versus $T$.  
 }
 \end{figure}    

Figure~\ref{Mix} shows the numerical simulations of a purely magnetic relaxation curve: $p$($t$) = 0.1$e^{- t}$+0.9$e^{- 6t}$ (MR: $W_M$ = 1), 
a mixed magnetic and quadrupolar relaxation curve: $p$($t$) = [ 0.1$e^{- t}$+0.9$e^{- 6t}$ ]$e^{- t}$ (Mix: $W_M$ = $W_Q$ = 1),
a stretched magnetic mixed relaxation curve: $p$($t$) = [ 0.1$e^{- \sqrt{t}}$+0.9$e^{- \sqrt{6t}}$ ]$e^{- t}$ (Mix1: $W_M$ = $W_Q$ = 1, $\beta$ = 0.5),
and a stretched quadrupolar mixed relaxation curve: $p$($t$) = [ 0.1$e^{- t}$+0.9$e^{- 6t}$ ]$e^{- \sqrt{t}}$ (Mix2: $W_M$ = $W_Q$ = 1, $\beta_Q$ = 0.5).
The stretched exponential function incorporates the short-$T_1$ distribution.  

We reanalyzed the recovery curve data in Fig.~\ref{L}(a) using Eq.~(\ref{sMQ}) with the fitting parameters $p$(0), $^{63}W_M$, $^{63}W_Q$, and $\beta$.   
The upper panel in Fig.~\ref{TT} shows $^{63}W_M$ and $^{63}W_Q$ versus temperature $T$ for the underdoped Hg1223 with $B_0 \parallel c$.
The lower panel in Fig.~\ref{TT} shows a variable exponent $\beta$ versus temperature $T$.
Above 160 K, $\beta$ close to 1 indicates a homogeneous relaxation.  
Below 140 K, $\beta$ slightly decreases on cooling. 
Then, the slightly larger values of $^{63}W_Q$ obtained using Eq.~(\ref{sMQ}) than those obtained using Eq.~(\ref{MQ}) were estimated.   
No significant effect of inhomogeneous relaxation is observed. 
The predominant magnetic relaxation rate and the very weak quadrupole relaxation rate obtained using Eq.~(\ref{sMQ}) were observed as well as those obtained using Eq.~(\ref{MQ}) for Hg1223.   

For the present Hg1223 samples, a separate evaluation of the inhomogeneous magnetic relaxation rate $W_M$ and the weak quadrupole relaxation rate $W_Q$ was achieved by using Eq.~(\ref{sMQ}) based on the exact solution.
The upper limit of $W_Q$ represents the values that cannot be quantitatively excluded.  

The finite $^{63}W_Q$ around 110 $< T <$ 140 K may be associated with low-temperature anomalies in the transport properties of underdoped Hg1223 at high external magnetic fields~\cite{MR}. 
The anomalies suggest a Fermi surface reconstruction with charge density wave ordering~\cite{MR}. 
The charge density fluctuations may induce the nuclear electric quadrupole relaxation immediately above $T_{\mathrm c}$. 
 
\section{Conclusions}
We found a simple reduced form of the theoretical recovery curve with both the magnetic and quadrupolar relaxation rates $W_M$ and $W_Q$ for the central transition line of the nuclear spin $I$ = 3/2 NMR. 
From the application of Laplace transformation, the reduced form of the exact solution was extended to include the inhomogeneous distributions of $W_M$ and $W_Q$.  
Using the reduced and stretched exponential forms,
we obtained the results of the predominant magnetic relaxation rate $^{}W_M$ and the upper limit of a very weak nuclear electric quadrupole relaxation rate $^{}W_Q$ for Hg1223. 

\end{document}